\documentclass[pra,twocolumn]{revtex4}
\usepackage{epsfig}
\usepackage{graphicx}
\usepackage{amsmath}
\usepackage{natbib}
\usepackage{color}
\newcommand{\n}{\nonumber}
\newcommand{\bn}{\begin{eqnarray}}
\newcommand{\en}{\end{eqnarray}}
\newcommand{\eml}{\end{multline}}
\newcommand{\bml}{\begin{multline}}
\newcommand{\h}{\hspace}

\newcommand{\op}[1]{\hat{#1}}

\begin{document}

\title {Rotation Sensitive Quench and Revival of Coherent Oscillations in a Ring Lattice}

 \author{Caelan Brooks$^1$, Allison Brattley$^1$ and Kunal K. Das$^{1,2}$}
 \affiliation{$^1$Department of Physical Sciences, Kutztown University of Pennsylvania, Kutztown, Pennsylvania 19530, USA}
\affiliation{$^2$Department of Physics and Astronomy, State University of New York, Stony Brook, New York 11794-3800, USA}
%\date{\today }
%
\begin{abstract}
We consider ultracold atoms trapped in a toroidal trap with an azimuthal lattice for utility as a macroscopic simulator of quantum optics phenomena. We examine the dynamics induced by the adiabatic introduction of the lattice that serves to couple the normal modes, as an analog of a laser field coupling electronic states. The system is found to display two distinct behaviors, manifest in the angular momentum - coherent oscillation and self-trapping -
reminiscent of non-linear dynamics, yet not requiring interatomic interactions. The choice is set by the interplay of discrete parameters, the specific initial mode and the periodicity of the lattice. However, rotation can cause continuous transition between the two regimes, causing periodic quenches and revivals in the oscillations as a function of the angular velocity. Curiously, the impact of rotation is determined entirely by the energy spectrum in the absence of the lattice, a feature that can be attributed to adiabaticity. We assess the effects of varying the lattice parameters, and consider applications in rotation sensing.
\end{abstract}
\maketitle

\section{Introduction}

Ultracold atoms trapped in ring shaped lattices offer a versatile system for probing and simulating quantum mechanics, while also offering possibilities for sensor applications \cite{Das-PRL-localization,Opatrny-Kolar-Das-rotation, Kwek-ring-lattice}. The closed loop topology enables access to some of the most enigmatic quantum features, like nonlocal effects \cite{Aharonov-Bohm, Thouless-book}, sustained superfluid flow \cite{Phillips_Campbell_superfluid_2013}, and effects of gauge fields mimicked by rotation \cite{Das-Christ,Das-PRL-localization}. The azimuthal lattice structure makes the system substantially richer, introducing band structure in a finite periodic system \cite{Guilleumas-nonlinear_ring}. Coherent states satisfying the periodic boundary condition also provide a macroscopic realization of de Broglie's classic model \cite{de-Broglie} of quantization of electronic orbits, that launched our current understanding of quantum physics. This last aspect of cold atoms in a ring lattice is the subject of interest in this paper.

\begin{figure}[t]
\centering
%\vspace{-0.2\linewidth}
\includegraphics[width=\columnwidth]{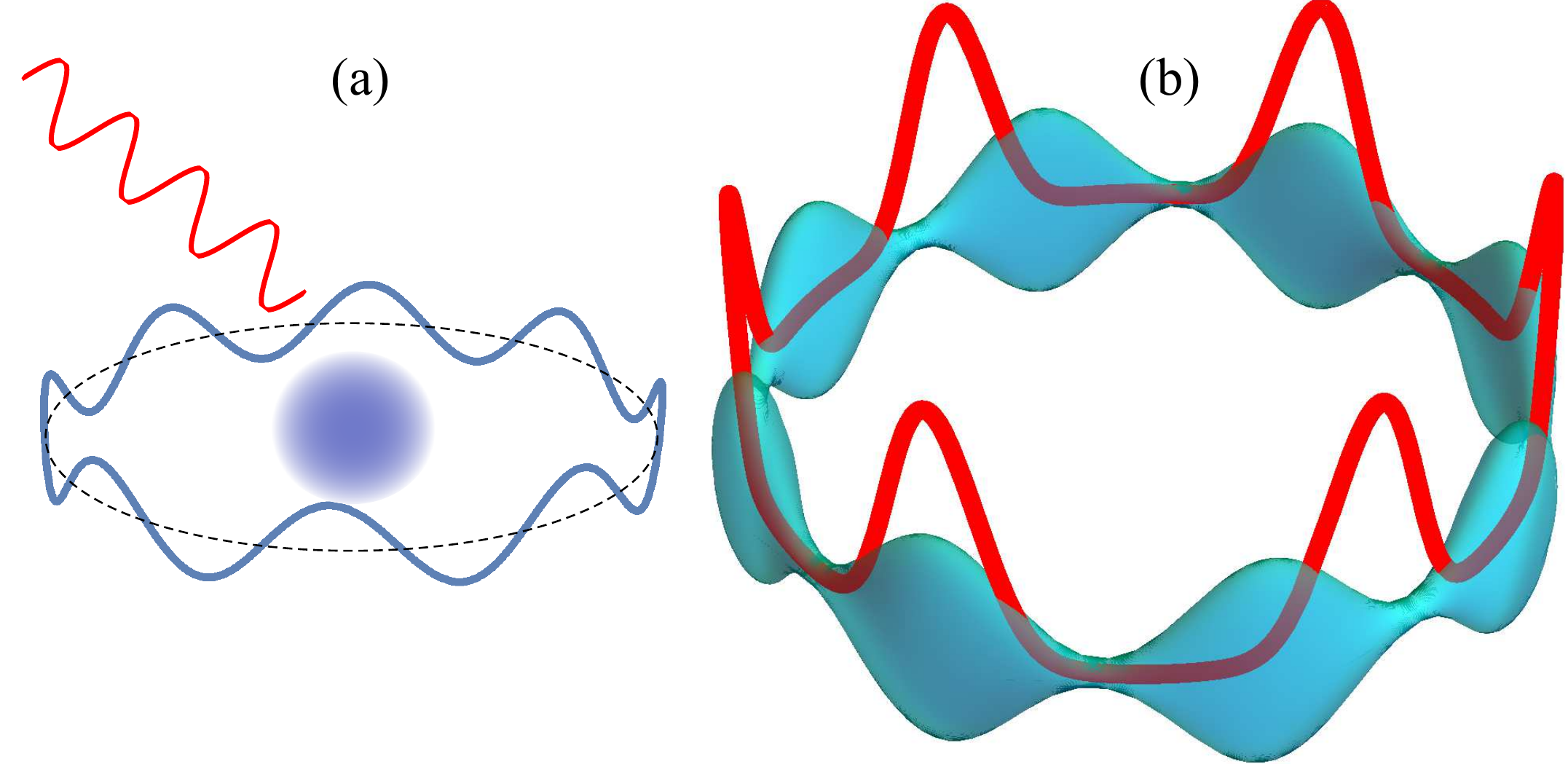}
\caption{(Color online) (a) Laser light interacting with an electronic state of an atom, its quantization represented by a de Broglie model of a eigen-mode satisfying periodic boundary condition. (b) This is mimicked by ulatracold atoms trapped in an effectively one-dimensional toroidal trap with an azimuthal lattice potential serving to couple the coherent collective modes of the atoms.}
\label{Fig1-Schematic}
\end{figure}

The introduction of a lattice serves to couple the normal modes of the ring. A direct analogy can be made with the situation of lasers interacting with electronic states of an atom \cite{Opatrny-Kolar-Das-rotation}, with the lattice coupling the modes of the ring just as a laser field couples the electronic states, illustrated schematically in Fig.~\ref{Fig1-Schematic}. Even spin degrees of freedom can be mimicked with clockwise and counter-clockwise flow orientations. This makes cold atoms in ring lattices an attractive simulator of quantum optics \cite{Mandel-Wolf} at a scale orders of magnitude larger. Beyond the similarities, the macroscopic scale, the external degrees of freedom, and the option of strong nonlinearity \cite{Bloch-RMP-Many-Body} create possibilities that can go beyond typical quantum optical paradigms, blending them with aspects of condensed matter physics and nonlinear dynamics.

Multiple pathways exist for creating ring traps for atoms \cite{ramanathan, Stamper-Kurn-2005,Riis-magnetic,Chapman-magnetic-ring-2001,Boshier-painted-potential,Foot-painted,Lee_Hill_photomask,
Mompart-conical,Dunlop-ring}, some conveniently adaptable to include an azimuthal lattice structure, such as the use of Laguerre-Gaussian (LG) beams \cite{Padgett, Zambrini:07}. While numerous experiments \cite{Campbell_resistive-flow,Phillips_Campbell_superfluid_2013,Phillips_Campbell_hysteresis} have been conducted with cold atoms in ring traps, proportionate effort with the inclusion of lattices are overdue, notwithstanding the rich physics indicated by continuing theoretical works \cite{Aghamalyan-AQUID, Aghamalyan-two-ring-lattice,Piza-ring-lattice,Tiesinga-soliton-lattice,Maik-dipolar,Moreno-Bose-Hubbard,Jezek-Bose-Hubbard-ring-lattice,
Jezek-winding-number,Doron-Cohen-1,Minguzzi-resonant-persistent,Minguzzi-PRA-two-bosons,Penna,Guilleumas-nonlinear_ring, Nigro_2018, Opatrny-Kolar-Das-LMG, Opatrny-Kolar-Das-rotation}.

The focus of  theoretical studies have been centered primarily on the nonlinear aspects of the system, and associated with persistent currents, and potential applications for atomtronics and quantum computation. These are very relevant topics and justify continued exploration. The purpose of this paper, in contrast, is to examine the linear dynamics of the system in the context of close analogies with quantum optics. We examine the dynamical behavior of ultracold atoms in a ring lattice, as a function of the primary lattice parameters with regards to two fundamental questions that are relevant for any experiment on such a system: (1) How does the system respond to the lattice being introduced adiabatically? (2) How does the presence of rotation impact the resulting dynamics? Our findings contain some surprising features that open lines of further studies and also provide principles that can find value in sensor applications.

In Sec. II, we present our physical model that is utilized through the paper. Section III identifies the salient features of the dynamics and explains them, and Secs.~IV and V clarify the influence of the parity of the initial mode and adiabaticity respectively. The crucial impact of rotation is presented in Sec.~VI with considerations of sensor application in Sec. VII. Effects of various lattice parameters on the induced dynamics are analyzed in Sec.~VIII. We conclude in Sec.~IX, with a discussion of feasibility in experiments and with an outlook for further studies.

\section{Physical Model}
We consider a BEC in a toroidal trap as shown in Fig.~\ref{Fig1-Schematic}. We take the minor radius to be much smaller than the major radius $R$ so that the system can be treated as a cylinder ${\bf r}=(z,r,\varphi)$ with periodic boundary condition on $z$.  We assume the confinement along $(r,\varphi)$, transverse to the ring circumference to be sufficiently strong to keep the atoms in the ground state
$\psi_r(r)\psi_{\varphi}(\varphi)$ for those degrees of freedom, so that the three-dimensional bosonic field operator can be written in the effective form $ \hat{\Psi}(z)\psi_r(r)\psi_{\varphi}(\varphi)$.
Integrating out the transverse degrees of freedom, the dynamics can be described by an effective one dimensional Hamiltonian
\begin{eqnarray}\label{QF-Hamiltonian}
&&\op{H}(t)=\int_0^{2\pi R}{\rm d}z\op{\Psi}^\dagger(z,t)\times\\
&&\left[- \frac{\hbar^2}{2m}\partial^2_z+V(z,t)+\frac{g}{4\pi l^2} \op{\Psi}^\dagger(z,t)\op{\Psi}(z,t)\right]\op{\Psi}(z,t).\n
\end{eqnarray}
where  $g=4\pi\hbar^2a/m$ is the interaction strength defined by the $s$-wave scattering length $a$, the mass of the atoms $m$, and the harmonic oscillator length $l$ for the transverse confinement along the cross-section of the torus. The potential along the ring is taken to be a periodic lattice, that can rotate with angular velocity $\omega$,
\begin{eqnarray}
V(z,t)&=&\hbar U(t) \cos^{2p}\left[{\textstyle \frac{1}{2}}q(z/R - \omega t)+\theta \right]\n\\&\equiv& \hbar U(t)\sum_{j=1}^{p}c_j \cos\left[j\{q(x - \omega t) +2\theta\}\right],\label{potential}
\end{eqnarray}
both $p$ and $q$ being positive integers. This represents a lattice potential of period $2\pi/q$ corresponding to $q$ peaks equally spaced along the circumference $2\pi R$ of the ring. Increasing the value of the power $2p$ of the cosine causes the peaks to become progressively narrower.  The second form of the potential introduces the scaled azimuthal co-ordinate $x=z/R$, and does a trigonometric expansion in terms of cosine functions of periods $2\pi/(jq)$ that are integer ($j$) divisors of the base period. The expansion coefficients $c_j=\frac{1}{2^{2p-1}}{2p\choose p-j}$ leave out a constant offset term $c_0=\frac{1}{2^{2p}}{2p\choose p}$. A phase shift $\theta$ is allowed for, but will be set to zero except where specified otherwise.

The strength $U(t)$  of the lattice potential is time dependent with the ability to turn the lattice  on and off with variable rates and duration. We will use the functional form,
\bn U(t)=U_0\times [\{1-e^{t^2/\tau^2}\}-e^{(T-t)^2/\tau^2}].\label{time-dependence}\en
The first pair of terms control the switching on, and the last term the switching off. The total duration of evolution is set by $T$ and the rate of switching on or off is controlled by $\tau$, with higher values causing a slower onset.

\begin{figure*}[t]\includegraphics[width=\textwidth]{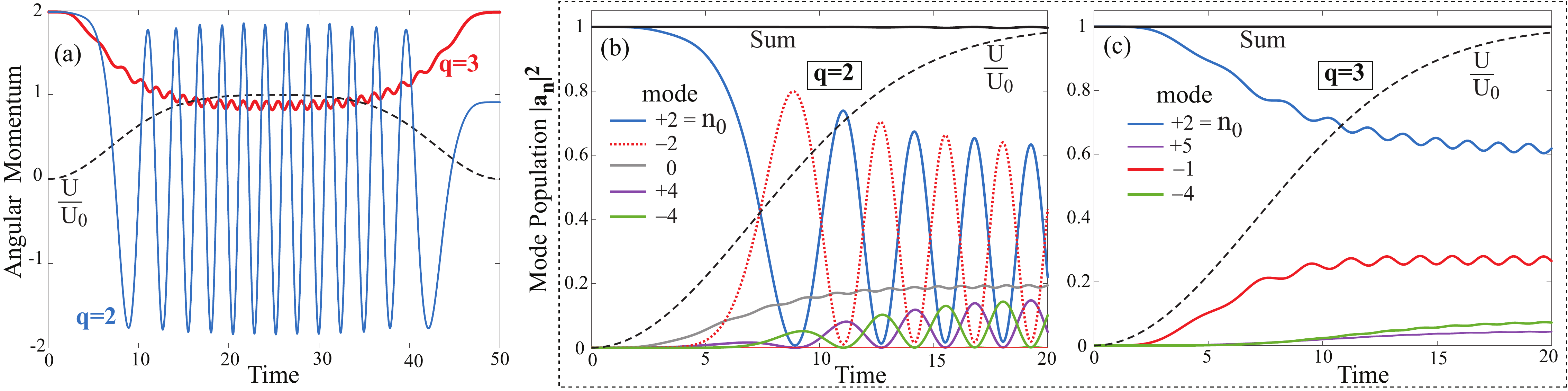}
\caption{(a) The time evolution of the net angular momentum (solid lines) of the quantum state in the ring is plotted as the lattice strength $U(t)$ (dashed black line) is switched on and off adiabatically. The thin blue line showing strong oscillation is for $q=2$ peaks in the lattice, and the thick red line with suppressed oscillation (self-trapped), is for  $q=3$. In both cases, the state is initiated in a specific circulating eigen-mode of the ring with quantum number $n_0=2$. The boxed panels (b,c) show the time evolution of the populations in significantly occupied modes, with (b) corresponding to the strongly oscillating angular momentum in panel (a) with $q=2$; and (c) corresponding to the self-trapped case in panel (a) with $q=3$, so called because the population persists significantly in the initial mode $n_0=2$. Only states with significant population are plotted, justified by their \emph{sum} remaining constant at unity during evolution, seen as a flat black line at the top.}\label{Fig2-angular-momentum}
\end{figure*}

Expanding the field operator in eigenstates of the ring
\bn \op{\Psi}(x)=\sum_{n=-\infty}^{\infty}\op{a}_{n}e^{-in\omega t} \psi_{n}(x);\h{5mm} \psi_{n}(x)=\frac{1}{\sqrt{2\pi}}e^{inx},\label{mode-expansion}\en
where $n=0,\pm 1, \pm 2, \cdots$  are the quantum numbers labeling the circulating eigen-modes. Our analysis of the dynamics in the ring will be based on how the quantum amplitudes of these modes evolve. The equation of motion for the Hamiltonian can be transformed into a series of coupled differential equations for the mode amplitudes or expansion coefficients $\op{a}_{n}$ \cite{Opatrny-Kolar-Das-rotation},
\bn i\frac{\partial}{\partial t}\op{a}_{n}(t)
&=&(\omega_n-n\omega)\op{a}_{n}+U(t)\sum_{j=1}^{p}c_j\left[\op{a}_{n-jq}+ \op{a}_{n+jq}\right]\n\\
&&+\frac{g}{4\pi^2 l^2} \sum_{k}\sum_{l}\op{a}^\dagger_{k} \op{a}_{l} \op{a}_{n+k-l}. \label{eqn_motion}\en
These equations serve to provide useful insight into the nature of the coupling between the modes as well as the impact of rotation: The lattice potential couples the modes in `ladders' of steps $\pm jq$ corresponding to the number of peaks $q$ in the lattice, with one such ladder for each available value of  $j=1,\cdots p$. Specifically, for the  lowest value $p=1$ in the exponent in Eq.~(\ref{potential}), only $j=1$ is available, so the modes coupled are separated by steps of $\pm q$. But for higher values, $p>1$, additional `ladders' with larger step sizes $\pm jq$ become available for every $j=2,\cdots p$. The angular velocity induced by rotation causes a shift of $n\omega$ in the spectrum, proportional to quantum number $n$ of each mode.

However, we will not use Eq.~(\ref{eqn_motion}) which, being a momentum space representation, is suitable for weak lattices, whereas in this study we will need to vary the lattice strength over a wide range. We will instead use the equation of motion in a position space representation, obtained directly from  the Hamiltonian Eq.~(\ref{QF-Hamiltonian})
\begin{eqnarray}
i\hbar\partial_t\op{\Psi}=\left[- \frac{\hbar^2}{2m}\partial^2_x+i\hbar\omega\partial_x+V+\frac{g}{4\pi^2 l^2} \op{\Psi}^\dagger\op{\Psi}\right]\op{\Psi}.
\label{eqn-motion-postion}
\end{eqnarray}
The field operator $\op{\Psi}(x,t)$ and the potential $V(x,t)$ have time and position dependence that are not displayed explicitly. The relevant connection between the two pictures represented by Eqs.~(\ref{eqn_motion}) and (\ref{eqn-motion-postion}) is via the projections $\hat{a}_n(t)=\langle\hat{\Psi}(x,t)|\psi_n(x)\rangle$. The effect of rotation has been added on with an angular momentum term which amounts to a transformation to a frame rotating with the lattice. This is consistent with the form of  Eq.~(\ref{eqn_motion}) obtained by explicit inclusion of the rotation induced phase in the mode expansion in Eq.~(\ref{mode-expansion}).

We will assume the mean field limit $\op{\Psi}\rightarrow\langle\op{\Psi}\rangle=\Psi$, applicable for a large number of particles when quantum fluctuations are relatively small. Correspondingly, we will represent mean field limit of the projection operators by $a_n(t)=\langle \hat{a}_n(t)\rangle$. We will also neglect nonlinearity, setting $g=0$, assuming weak interactions possible by Feshbach resonance \cite{Tiesinga-RMP}. Along with length unit $R$, our units assume $\hbar=m=1$.  We will examine the coherent dynamics of the quantum state of the medium in the ring, as it interacts with the lattice, by computing the projections $a_n(t)$ that give the amplitude of the modes, with $|a_n(t)|^2$ tracking their population.  We will refer to the modes $\psi_n$ simply by their quantum number $n$.

\section{Criteria for Coherent Oscillation}\label{Sec:Criteria}

The eigenstate expansion Eq.~(\ref{mode-expansion}) represents circulating modes, with positive and negative values of the quantum number corresponding to counter-propagating flows.  We initiate the system in one of the modes with quantum number $n_0$, with the lattice potential switched off, $U(t)=0$. Then, we adiabatically switch on the lattice to reach a stable maximum depth of $U_0$. As the equations of motion Eq.~(\ref{eqn_motion}) show, this causes the state to couple to other modes with quantum numbers $\{n_0\pm q,n_0\pm 2q,\cdots\}$ transferring population.  It is clear from the properties of the binomial series, that for a fixed value of $p$, the expansion coefficients $c_j$ used in Eq.~(\ref{potential}) diminish with increasing $j$ so that the highest weight will always correspond to the first term with $j=1$, which therefore provides the dominant coupling pathway. Therefore, we first consider the case of the lowest power with $p=1$ which contains only this single term in the expansion. The effects of higher powers, $p>1$ will be examined in Sec.~\ref{Sec:lattice-param}.

We solve the differential equation, Eq.~(\ref{eqn-motion-postion}) with $g=0$, for the time-evolved quantum state  $\Psi(t)$, and from its projections $a_n(t)$, we determine the populations of a set of modes $n_0\pm jq, j=0,1,\cdots$ on a ladder centered about the initial mode $n_0$.  Energy conservation places a natural limit on the number of such modes that will acquire significant population. We ensure that in our simulations, a sufficient number are taken into consideration by verifying the sum of their populations remains unity.

We first capture the aggregate behavior of the medium by computing the expectation of the angular momentum of the system $\langle i\hbar \partial_x\rangle=\langle \Psi(t)|i\hbar \partial_x|\Psi(t)\rangle$. Two very different and distinct behavior emerges as shown in Fig.~\ref{Fig2-angular-momentum}:  (I) The angular momentum undergoes oscillations with almost complete reversal occurring periodically, or (II) The angular momentum stabilizes at some intermediate value with the emergence of small oscillations about that mean, that we will refer to as `wiggles'. To contrast with the oscillating case, we refer to the second case as `self-trapped' in analogy with similar behavior in non-linear dynamics, wherein the population is restrained from making complete transition out of a specific state.

By adiabatically switching off the lattice, we find that for the self-trapped cases, the angular momentum is restored to the initial value as illustrated in Fig.~\ref{Fig2-angular-momentum}. However in the oscillating case, the degree of restoration depends on the precise timing of initiating the switching off, but complete revival appears to be always possible.

\begin{figure}[t]
\centering
%\vspace{-0.2\linewidth}
\includegraphics[width=\columnwidth]{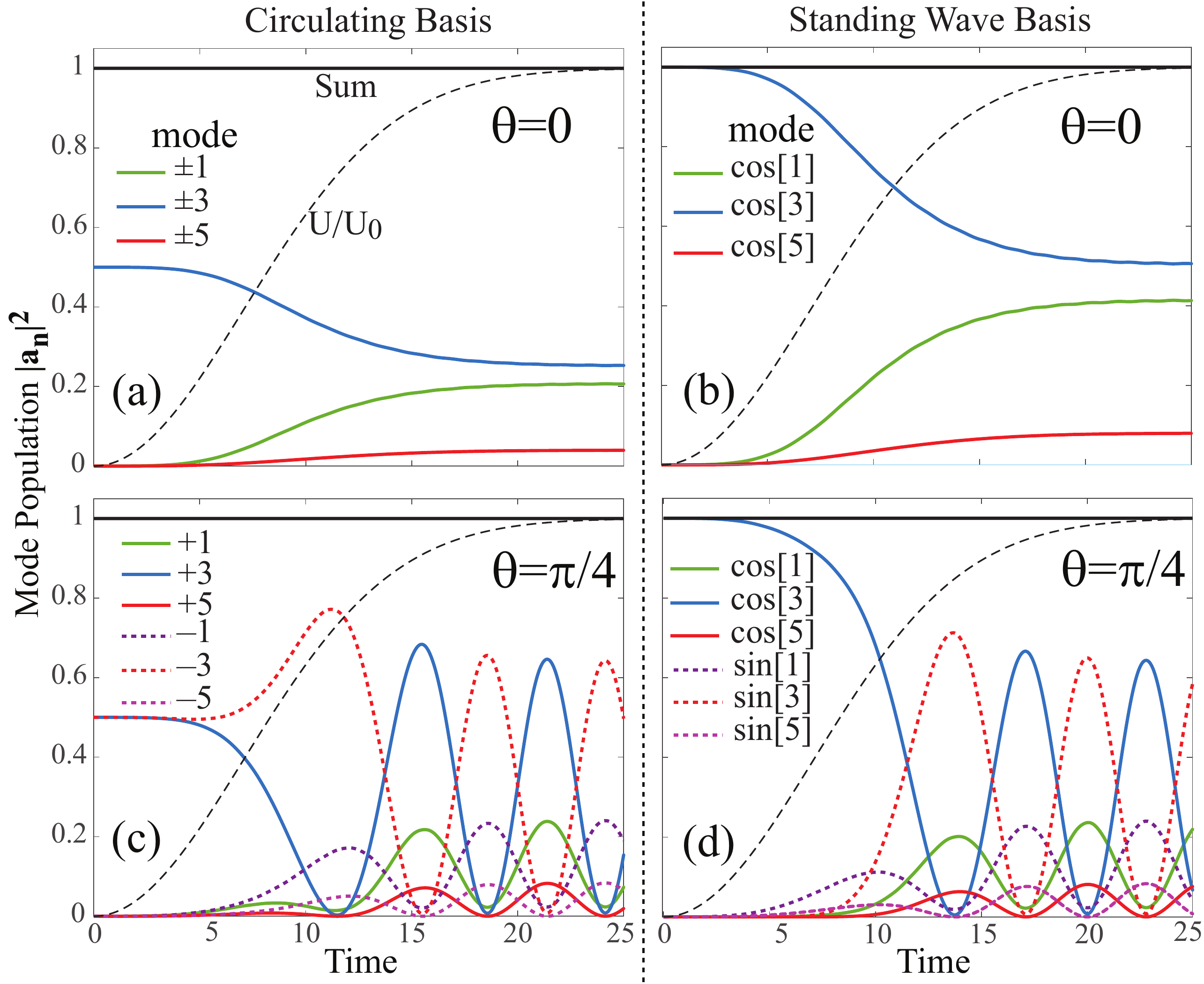}
\caption{(Color online)  When the initial mode is a standing wave $\cos(n_0x)$, the time evolution of the population in occupied modes is plotted as the lattice potential $U(t)$ (dashed black line) is switched on adiabatically, shown here for $n_0=3$ and $q=2$. Left panels (a,c) show the circulating mode basis, and the right panels (b,d) the standing wave basis (labeled $\cos(nx)\equiv \cos[n]$).  In both bases, upper panels (a,b) show that with no phase shift, $\theta=0$, between the lattice and the initial mode, oscillations are suppressed. But in the lower panels (c,d) with $\theta=\pi/4$, oscillations occur. Note, in panel (a) counter-propagating  modes with the same wave number ($\pm n$) have identical evolution, and overlap. The states displayed are the only ones with significant population, confirmed by their sum (flat black line) remaining normalized.}
\label{Fig3-Standing}
\end{figure}

Whether oscillations occur or not is determined by the relation between the spatial periodicities of the initial mode, $n_0$, and that of the lattice, $2\pi/q$. Oscillating cases constitute a limited subset of possibilities that occur whenever $2n_0/q=j$ is a non-zero integer, also necessitating that $q\leq |2n_0|$. Examination of the evolution of the population sheds light on this: Satisfaction of this condition provides a ladder pathway through a sequence of intermediate coupled modes from the initial mode $n_0$ to another mode which is degenerate in energy with it, in this case $ -n_0$. Thus, in Fig.~\ref{Fig2-angular-momentum}(b), where $q=2$ the population dominantly oscillates between the degenerate modes $n_0=+2$ and $n_0-2q=-2$, because they are coupled by two ladder steps of $q$. The intermediate state $n_0-q=0$ is populated but has muted oscillations since it is non-degenerate. Population oscillations also occur between other degenerate states, for example between the modes $n=\pm 4$, since there is direct coupling of one of them with the initially occupied mode: $n_0+q=+4$. But the amplitude is lower than for oscillations involving the initial mode directly, due to larger energy gap.

In contrast, for the self-trapped cases, the ladder of coupled modes does not contain any pair with degenerate energies as illustrated in Fig.~\ref{Fig2-angular-momentum}(c). There is however some exodus of population from the initial mode, and as should be expected, the primary recipient is the mode closest to it on the ladder but with lower energy. There is progressively less population appearing in modes energetically farther apart on the ladder. An interesting special case of note is that of the ground state.  Since it is non-degenerate, it always remains self-trapped regardless of the periodicity of the lattice.

\section{Parity Effect for Standing Waves}

The presence of oscillations is sensitive to the initial state. If we initiate the system in a standing wave of form $\sim \cos(n_0x)$, there are never any instances of oscillating cases even when the criteria defined in Sec.~\ref{Sec:Criteria} are satisfied. That is because the potential $V\sim\cos(qx)$ is an even function of the co-ordinate, and therefore only couples states with the same parity which, in this case, are other standing waves also of the form $\cos(nx)$ with $n=0,1,2,\cdots$. Since $n\neq n_0$, there can be no degeneracies among those states.  Strong oscillations can only occur between a pair of distinct states that are degenerate in eigenenergies \emph{and} are also separated by an integer number of ladder steps $\pm q$, set by the lattice period. This requires coupling of $\cos(n_0x)$ with $\sin(n_0x)$. In order to make that happen, we introduce a phase shift of $\theta=\pi/4$ into the potential, so that $V\sim\cos[qx+2\theta]\rightarrow \sin(2qx)$ an odd parity function that can now couple even and odd parity states. Physically, in the context of $p=1$, this amounts to initiating the system in a mode that is phase-shifted by $\pi/2$ from the period of the lattice peaks. Figure~\ref{Fig3-Standing} shows the case where the initial mode is a standing wave $\cos(3x)$, in a lattice of periodicity $q=2$.  Oscillations are absent with $\theta=0$ but appear with $\theta=\pi/4$.

The criteria for oscillation is now different. The ladder of coupled modes must necessarily contain a pair of modes that are resonantly coupled which can only happen if modes with quantum number $n=\pm q/2$ (set by the lattice period) are on the ladder, requiring $q$ to be even. That ensures that the degenerate modes $\cos(qx/2)$ and $\sin(qx/2)$ are directly coupled.  Then, there exists a pathway for the initial mode $\cos(n_0x)$ to couple with its energy-degenerate pair $\sin(n_0x)$. This can only happen if $2n_0/q$ is an \emph{odd} integer, otherwise any ladder connecting $n_0$ to $-n_0$ would lead to the non-degenerate zero mode and there can be no resonant coupling.

Two different bases are used to show the dynamics of the states, with Fig.~\ref{Fig3-Standing}(a,c) showing the population in circulating modes, and Fig.~\ref{Fig3-Standing}(b,d) showing that for standing wave modes. In the standing wave modes, only cosine modes populate as expected for $\theta=0$, but oscillations with sine mode appear when $\theta=\pi/4$.  In the circulating bases, both clockwise and counter-clockwise modes have exactly the same population evolution for identical mode number, when $\theta=0$. But, with $\theta=\pi/4$, when oscillations occur, the oppositely circulating modes evolve differently, demonstrating consistency regardless of the specific basis used.

One feature we have noted across all strongly oscillating cases, regardless of the initial mode - circulating or standing wave - is that at steady state, the clockwise modes all fall into sync with each other and the counter-clockwise modes likewise sync up mutually. This can be clearly seen in Fig.~\ref{Fig3-Standing}(c), where the positive numbered modes all oscillate together in phase and so do all the negative numbered modes. In contrast, there is no such syncing of cosines and sines evident in Fig.~\ref{Fig3-Standing}(d) rather the cosine modes are out of phase with the adjacent cosine modes.  This kind of resonant directional `mode-locking' behavior ensures strong oscillation of the angular momentum, when coherent oscillations are present.

\section{Instantaneous Eigenstates}

The dynamics as described assumes adiabatic switching of the potential. The implications can be understood by following the projection of the wavefunction on the instantaneous eigenstates of the Hamiltonian as it evolves, parameterized by the time $t$.  With sufficiently slow evolution, the system will follow the instantaneous eigenstate $\phi_{n0}(t)$ that evolved from the mode $\psi_{n0}$ the system was initiated in.  Being a closed system, in the absence of rotation all the instantaneous states can be chosen to be real valued, that take the form $\sin(nx)$ and $\cos(nx)$ in the limit of no lattice. As the lattice is switched on, their functional form alters, but they can still be classified as symmetric or antisymmetric about the potential maxima, designated $\phi_{nS}$ and $\phi_{nA}$ respectively. When the initial mode is a circulating one, the instantaneous state $\phi_n$ is a $50/50$ superposition of $\phi_{nS}$ and $\phi_{nA}$.

\begin{figure}[t]
\centering
%\vspace{-0.2\linewidth}
\includegraphics[width=\columnwidth]{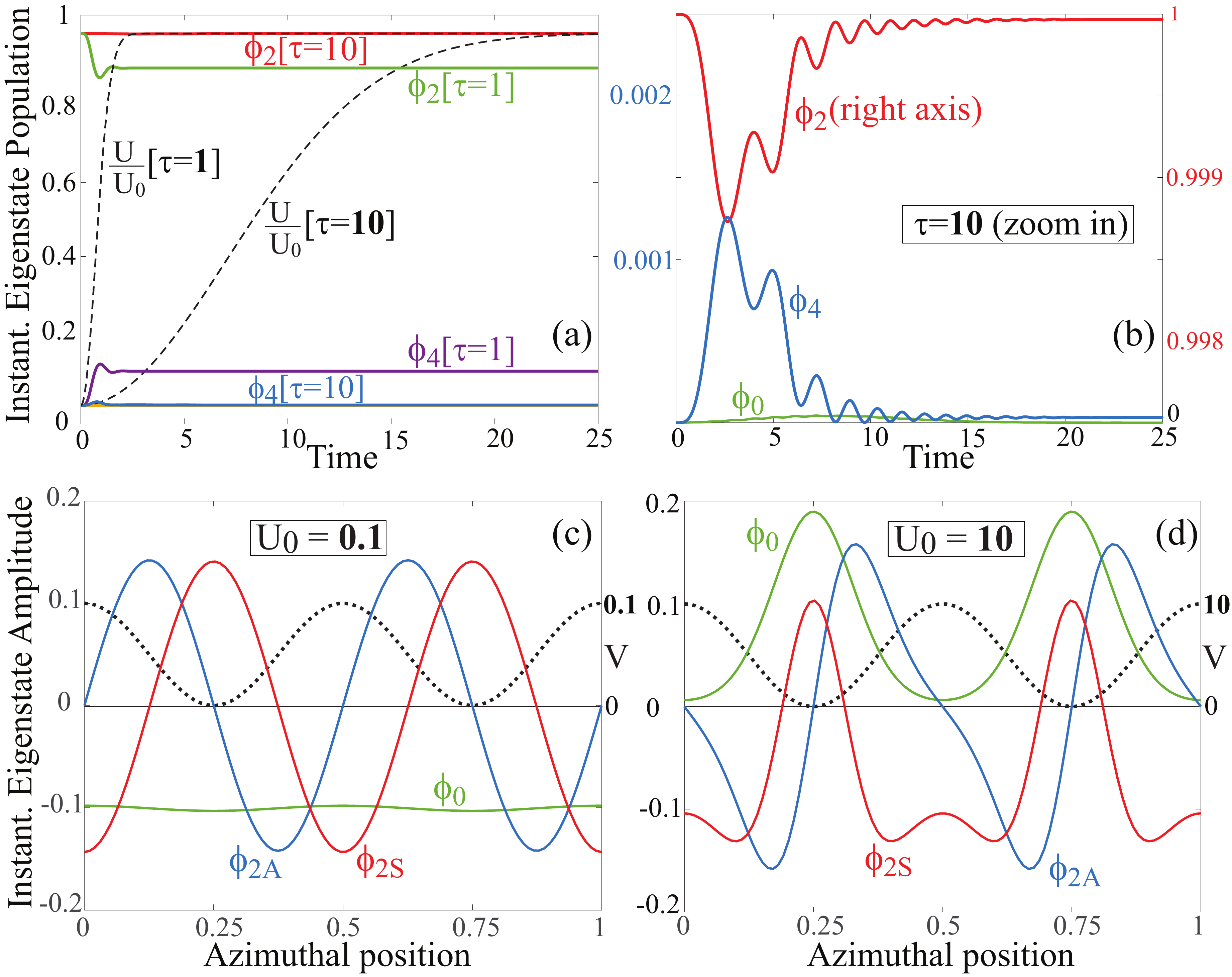}
\caption{(Color online)(a) Evolution of the population projected on the \emph{instantaneous} eigenstates $\phi_n$ of the Hamiltonian, for $q=2$ and initial mode $n_0=2$. The solid lines show the population in the occupied states , shown for two different rates (set by $\tau$) of switching on of the potential $U(t)$ plotted in dashed black lines. Faster rate ($\tau=1$) causes visible loss from the initial mode $\phi_2$, but hardly any at the slower rate ($\tau=10$). (b) Zooming in for ($\tau=10$) shows any loss is sufficiently small to justify adiabaticity. (c,d) Plots of the instantaneous eigenstates that are populated are shown: (c) when the lattice potential $V$ is weak, $U_0=0.1$ and (d) when it is strong $U_0=10$. The states are labeled symmetric (S) or antisymmetric (A) about the peaks of the lattice potential $V$ plotted in dotted black line on right axis.}\label{Fig4-adiabatic}
\end{figure}

In our simulations, the lattice potential is switched on or off according to Eq.~(\ref{time-dependence}), where a larger value of the parameter $\tau$ causes a more adiabatic variation. Figure~\ref{Fig4-adiabatic}(a,b) shows that for $\tau=1$, the instantaneous state $\phi_{2}$ loses some population during the switching on, but then stabilizes.  For a more adiabatic onset of the potential, with $\tau=10$, there is hardly any loss of population, only detectable on zooming in at the order of  $10^{-3}$.

The composition of the instantaneous states reflects the oscillations present in the normal modes of the ring without a lattice. Figure~\ref{Fig4-adiabatic}(c,d) show the instantaneous eigenstates corresponding to initial mode $n_0=2$ at weak and strong lattice strengths. At weak lattice, the relevant states are sinusoidal, symmetric and antisymmetric about the two peaks of the lattice with $q=2$, $\phi_{2A}\simeq \sin(2x)$ and $\phi_{2S}\simeq \cos(2x)$. As the lattice reaches maximum depth,  $\phi_{nA}\simeq 0.44\sin(2x)+0.064\sin(4x)+0.0015\sin(6x)$  and $\phi_{2S}\simeq 0.44+0.21\cos(2x)+0.092\cos(4x)+0.0030\cos(6x)$, reflecting relative populations seen in Fig.~\ref{Fig2-angular-momentum}(b).

The antisymmetric state here actually has the lower energy, which may seem surprising considering that antisymmetric pairing of localized states always has higher energy. The reason is that in this case of two lattice sites, $q=2$, as the lattice is turned on, the ground state remains symmetric, pairing up with the antisymmetric state arising from the initially degenerate pair of first excited states. The two states approach a new degeneracy as the lattice strength is increased. Likewise the symmetric first excited state pairs up with the antisymmetric second excited state, and so on, maintaining the expected pattern of antisymmetric state having the higher energy within each pairing, as seen later in Fig.~\ref{Fig6-energy}(d).

When we initiate the state in the circulating mode $\psi_2$ in the absence of a lattice, it comprises of two degenerate modes, of which the antisymmetric one migrates to the lower band and the symmetric to the higher band, causing the seeming inversion of the energy structure of symmetric/antisymmetric instantaneous states when a correspondence is made with the initial modes $\psi_n$.

\section{Sensitivity to Rotation}

\begin{figure}[t]
\centering
%\vspace{-0.2\linewidth}
\includegraphics[width=\columnwidth]{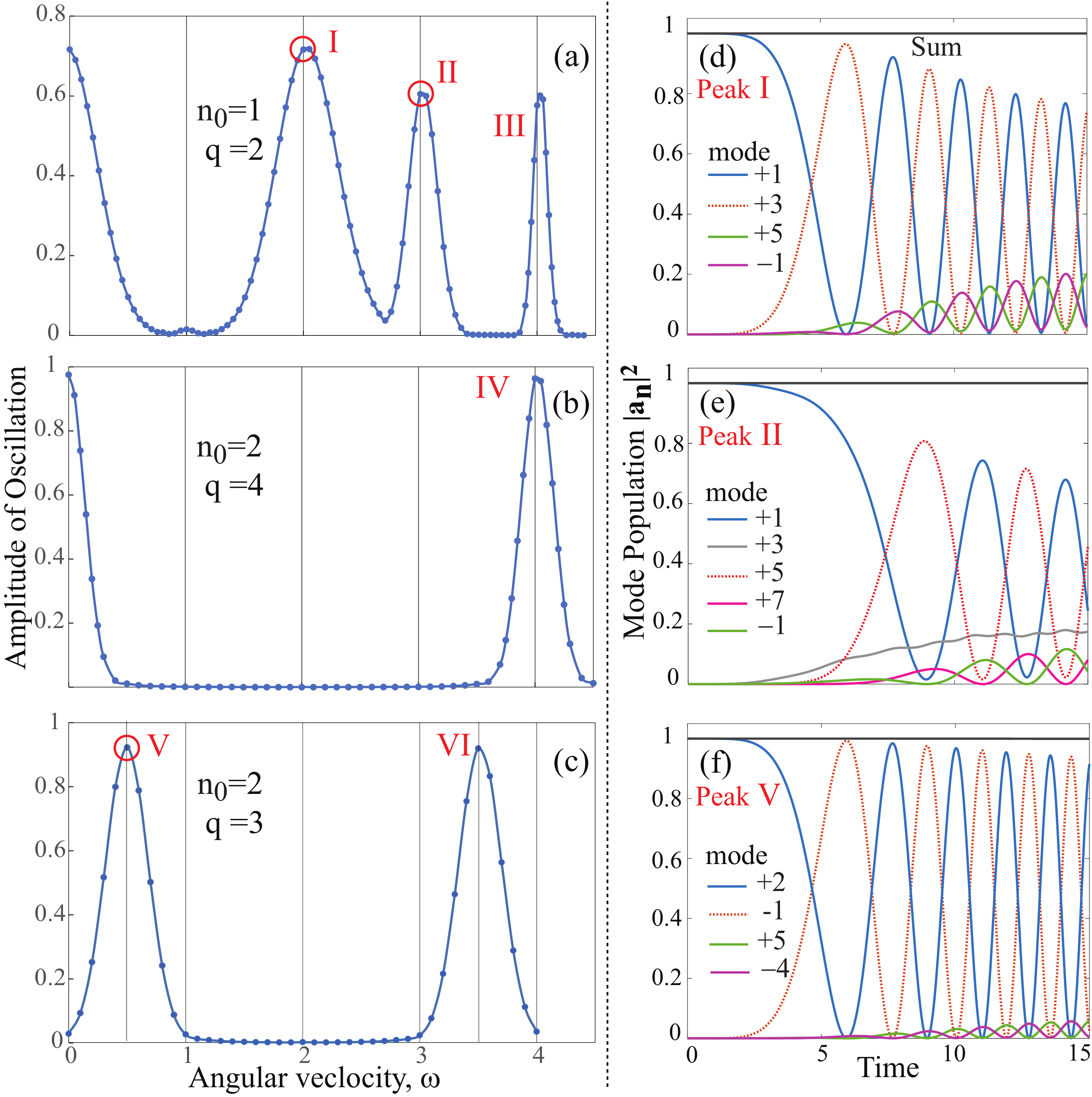}
\caption{(Color online) (a,b,c) Rotation changes the dynamics continuously between oscillating and self-trapped behavior, evident in the variation of the amplitude of oscillation of the initial state $n_0$. At no rotation $\omega=0$, cases in panels (a,b) are oscillatory and in panel (c) self-trapped, but all show recurrence of oscillatory behavior at specific values of the angular velocity $\omega$.  (d,e,f) The right panels display the time evolution of the modal population for parameters corresponding to specific peaks labeled by Roman numerals on the left panels. Specifically, panels (d) and (e) correspond to peaks I and II of panel (a): Note the initial mode $n_0=1$ dominantly couples to mode $n=3$ in (d)  but with  mode $n=5$ in (e) showing that rotation alters the primary coupling.}\label{Fig5-recurrence}
\end{figure}

Introducing rotation fundamentally changes the dynamics so far discussed. Rotation can transform an oscillating case into a self-trapped case and vice versa as shown in Fig.~\ref{Fig5-recurrence}: In its panels (a) and (b), for an initially oscillatory case, as the angular velocity is increased, the amplitude of oscillation of the initial mode diminishes and practically vanishes. Similar behavior is seen with the the mean value, together indicating a transition to self-trapped behavior. Curiously, at larger angular velocities, the oscillations undergo an almost complete revival, that repeats at integer values of the angular velocity.

Likewise, for an initially self-trapped case, increasing angular velocity causes transformation into an oscillatory case, as seen in Fig.~\ref{Fig5-recurrence}(c). There are once again revivals of oscillation which now peak at \emph{half} integer values of the angular velocity.

The population evolution at the angular velocities where the revival peaks occur reveal another aspect in Fig.~\ref{Fig5-recurrence}(d,e,f). The dominantly coupled modes are different from the ones when there is no rotation. Thus, for initial mode, $n_0=1$ and lattice periodicity $q=2$, with no rotation $\omega=0$, the population oscillates between $n=\pm 1$, but at $\omega=2$, the dominant oscillation occurs between $n=1$ and $n=3$ as seen in panel (d) while at $\omega=3$ it occurs between $n=1$ and $n=5$.  Similar drift of the dominantly coupled modes is observed to occur when the system is self trapped in the absence of rotation, with the mode oscillations corresponding to the first revival peak in Fig.~\ref{Fig5-recurrence}(c) shown in Fig.~\ref{Fig5-recurrence}(f).

\begin{figure}[t]
\centering
%\vspace{-0.2\linewidth}
\includegraphics[width=\columnwidth]{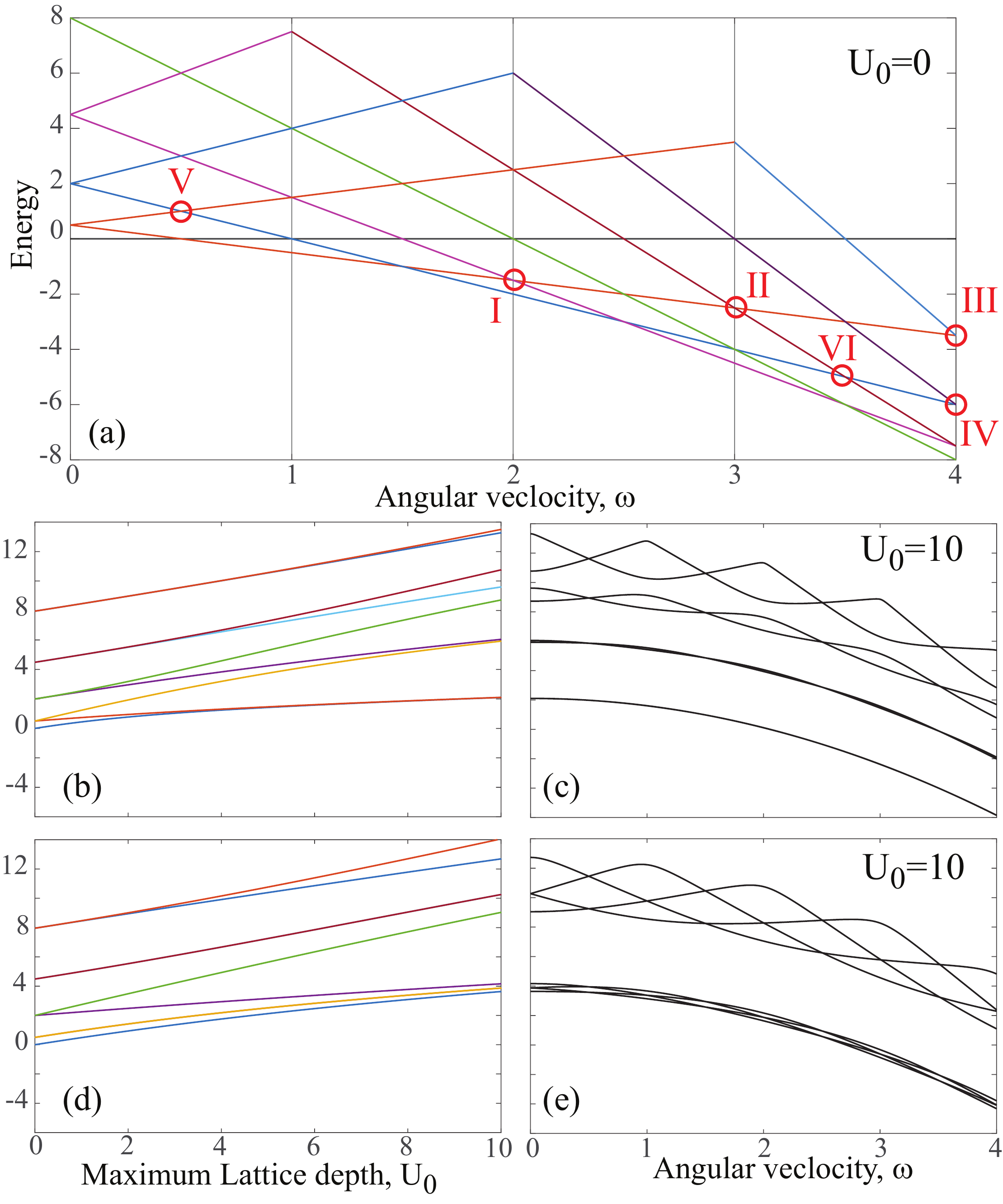}
\caption{(Color online) Variation of the lowest energy eigenvalues in the ring as function of: (a) the angular velocity with no lattice, $U_0=0$; (b,d) the lattice strength without rotation $\omega=0$; (c,e) the angular velocity for lattice strength $U_0=10$. Panels (b,c) are for $q=2$ and (c,e) for $q=4$, reflected in the number of states in each band when the lattice is on. Intersections labeled by Roman numerals in the top panel match similarly labeled peaks in Fig.~\ref{Fig5-recurrence} that mark recurrence of oscillation. They correspond to the onset of new degeneracies for the initial mode, induced by rotation, with the lattice absent. The four lower panels show that those relevant degeneracies are conspicuously absent with the lattice present.}\label{Fig6-energy}
\end{figure}

An explanation for this behavior can be found in the energy spectrum of the ring in the presence of rotation shown in Fig.~\ref{Fig6-energy}(a).  The oscillation revival peaks correspond to the emergence of new degeneracies. Considering that in the absence of a lattice the modes have energy $E_i=n_i^2/2-n_i\omega$ (setting $\hbar=m=1$), degeneracies of two modes $E_i=E_j$ occur at $\omega=(n_i+n_j)/2$, that takes integer and half integer values. The $n_0=1$ mode establishes new degeneracies with $n=+3$ and $n=+5$ at $\omega=2$ and $\omega=3$, and those are precisely the dominant modes in Fig.~\ref{Fig5-recurrence}(d,e) that correspond to the peaks I and II in the amplitude at those $\omega$ in Fig.~\ref{Fig5-recurrence}(a). All of the amplitude peaks in that figure are similarly explained, including the initially self-trapped case in Fig.~\ref{Fig5-recurrence}(c). Each peak corresponds to the emergence of a new relevant degeneracy for the initially occupied mode, as indicated in Fig.~\ref{Fig6-energy}(a).

Notably, we also see why there is no peak in Fig.~\ref{Fig5-recurrence}(a) at $\omega=1$, because Fig.~\ref{Fig6-energy}(a) shows that the $n_0=1$ mode is not degenerate with any other mode, and hence oscillations are suppressed. However, energy degeneracy is not the complete story. For example in panel Fig.~\ref{Fig5-recurrence}(b) there are no peaks at $\omega=1$ and $3$ or for panel Fig.~\ref{Fig5-recurrence}(c) at $\omega=1.5$ and $2.5$, although Fig.~\ref{Fig6-energy}(a) shows that the initial mode $n_0=2$ has degeneracies there. That is because, as explained in Sec.~\ref{Sec:Criteria}, for oscillations to occur there also needs to be a ladder pathway between the degenerate modes, that is, they need to be separated by integer multiples of the lattice period $q$, which does not occur for these above mentioned cases.

Perhaps the most surprising aspect of these quenches and revivals of the coherent oscillation is that they are completely determined by the energy spectrum in the \emph{absence} of the lattice.  Figure~\ref{Fig6-energy}(b,d) show the variation of the spectrum as $U_0$ is varied from zero to the maximum value for lattice period $q=2$ and $4$, as used in Fig.~\ref{Fig5-recurrence}, and then Fig.~\ref{Fig6-energy}(c,e) show how the spectrum is further changed by angular velocity, keeping the lattice strength  at $U_0=10$. Clearly, the defining degeneracies that appear in the absence of the lattice in  Fig.~\ref{Fig6-energy}(a) are generally absent in Fig.~\ref{Fig6-energy}(c,e). This underscores the relevance of the adiabatic switching on of the potential ($\tau=10$ in Eq.~\ref{time-dependence}) with the angular velocity already present. As shown in Fig.~\ref{Fig4-adiabatic} earlier, the state remains in the instantaneous state as the lattice is turned on, so that the degeneracies present in the absence of the lattice continue to determine the dynamics even when the lattice reaches maximum strength, as long as it is done so adibatically.

\section{Sensor Application Potential}

\begin{figure}[t]
\centering
%\vspace{-0.2\linewidth}
\includegraphics[width=\columnwidth]{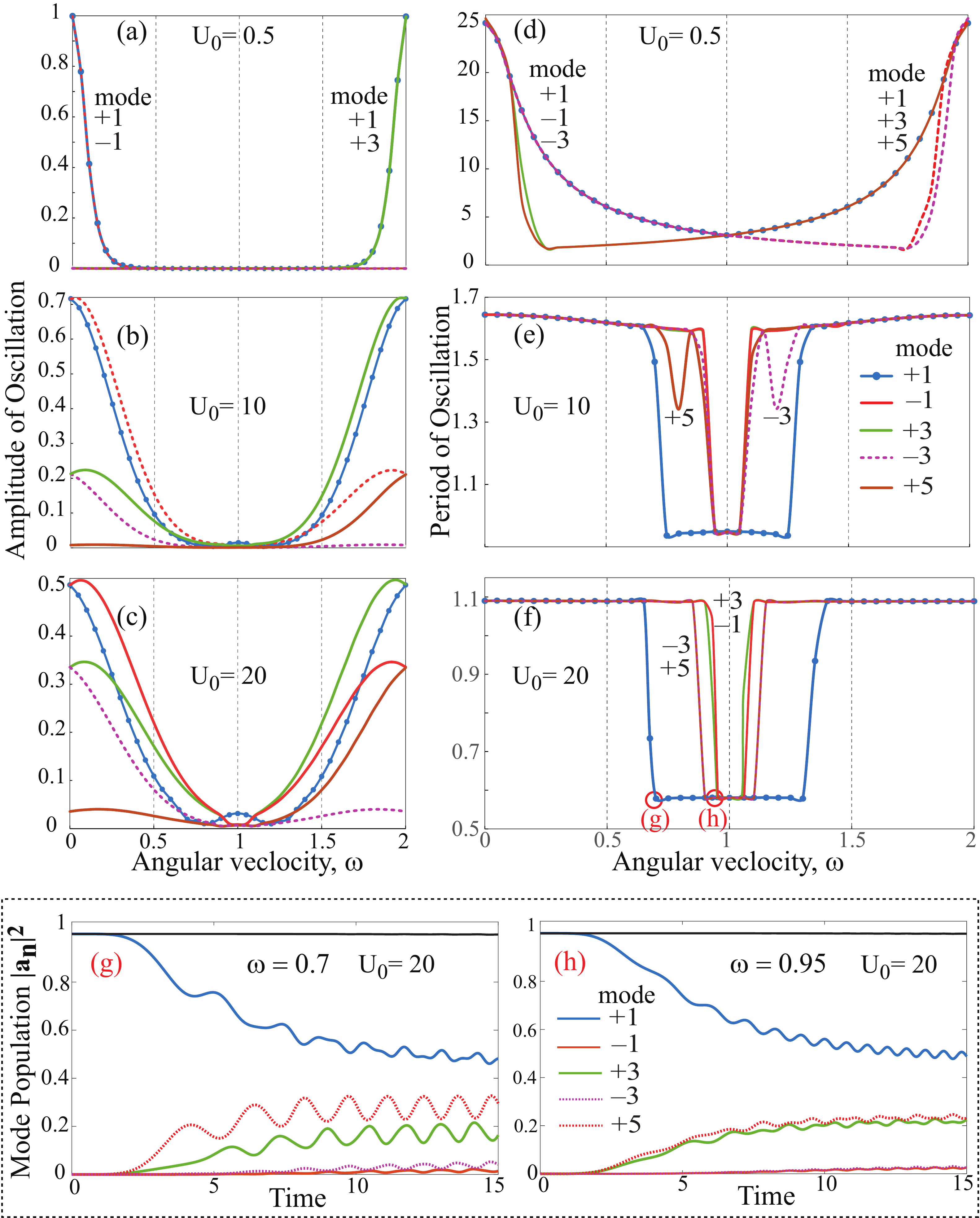}
\caption{(Color online) \emph{The legend in subplot (e) applies to subplots (a-f).}  The \emph{left panels} (a-c) plot the variation of amplitude of oscillation caused by rotation. They match Fig.~\ref{Fig5-recurrence}(a) except here, along with the initial mode $n_0=1$ (marked by blue dots), the modes it couples to are also shown to undergo similar behavior, and all are impacted by the lattice strength $U_0$. The \emph{right panels} (d-f) show that rotation also impacts the period of oscillation of the modes causing a drop and subsequent rise, that gets sharper with increasing $U_0$.  Panels (a,d) label modes that overlap with the initial mode; its dominant coupling changes from mode $n=-1$ to $n=+3$ as $\omega$ varies, also seen at higher $U_0$ but less stark.  \emph{Boxed panels} (g,h) show time evolution of the primarily occupied modes, corresponding to the specific values of $\omega$ \emph{circled} in panel (f); fluctuations are due to migration of the dominant coupling.}
\label{Fig7-rotation}
\end{figure}

The sensitivity of the coherent oscillation to rotation naturally suggests possible application in rotation sensing.  The response of the amplitude of oscillations will be manifest in the angular momentum, hence the circulation, an observable of the system. However, Fig.~\ref{Fig5-recurrence} shows that the decline in the amplitude with rotation is rather gradual, indicating limits to sensitivity. But Fig.~\ref{Fig7-rotation}(a,b,c) indicates this can be remedied by reducing the lattice depth. In the limit of very weak coupling (small $U_0$), the drop in amplitude of oscillation can be quite sharp, indicating heightened rotation sensitivity. Furthermore, as seen there, weak lattice also has the advantage that only the primary modes are excited, making the dynamics more transparent and cleaner.

There is of course a trade-off. Weaker lattice leads to longer periods, evident in Fig.~\ref{Fig7-rotation}(d,e,f), which reflects the energy time trade-off of the uncertainty principle. So that while rotation sensitivity can be enhanced by weakening the lattice strength, longer duration of observation is required to allow for completion of a period.

Interestingly, a potential bypass of this trade-off also seems to be implied in the behavior of the period of oscillation.  At some non-zero angular velocity, there is a sharp drop in the period as the angular velocity is increased. This is shown in Fig.~\ref{Fig7-rotation}(d,e,f).  That drop is sharper at higher lattice strengths. But, the trade-off here is that the drop off is smaller in magnitude.  One could, in principle, utilize the sharpness of period change, if one were to monitor the period of oscillation keeping the system rotating at a uniform rate close to the transition point. Then, a small change in the period would indicate a change in the rotational state.

The drop in both the period, as well as the amplitude of oscillation and subsequent revival, actually occurs for all the relevant coupled modes, although to a variable degree for each. This marks the transition of one set of strongly coupled oscillating modes to another set, as the energy degeneracy shifts with rotation. This is clearly evident in Fig.~\ref{Fig7-rotation}(a,b,c) for the amplitudes of the modes as $\omega$ changes.

The regime of lowered period span different ranges of the angular velocity for the various modes, but is widest for the initial mode as seen in Fig.~\ref{Fig7-rotation}(d,e,f). The cause for the sudden change is evident when one tracks the population oscillation for values of $\omega$ lying in the valley of lowered periodicity. For a pair of points in that regime, Fig.~\ref{Fig7-rotation}(g,h) shows that oscillations become rather erratic.  This can be be attributed to transitions in the coupling of the modes that occur here, so that the same mode is coupled with equal strength to two distinct modes.

\section{Influence of Lattice parameters}\label{Sec:lattice-param}

We now turn our attention to the influence of the lattice parameters on the system dynamics. To be specific, we track the response of the time evolution of the population in the initial state, characterizing it by the period and the amplitude of its oscillation, as well as by the mean value of the population. All are averaged over multiple periods at steady state when the lattice potential is constant at maximum depth. We illustrate this comparatively for an oscillating case $n_0=2,q=2$ and a self-trapped case $n_0=2, q=3$. In the latter case, the period and the amplitude refer to that of the wiggles in the population of the initial state.

\subsection{Revival Time}

As noted earlier, on adiabatically switching off the lattice, in the oscillating cases, the system does not necessarily restore to the initial state and angular momentum. This is influenced by the rate of switching off as well as by when the switch off is initiated.  By adjusting these parameters, via $\tau$ and $T$ in Eq.~(\ref{time-dependence}), practically any terminal angular momentum can be achieved intermediate between and upto complete circulation clockwise or counter-clockwise. The time for complete revival depends on the number of ladder steps, given by the ratio $2n_0/q$, separating the initial state from its energy degenerate counterpart. A higher value requires longer revival time. Furthermore, for fixed lattice period $q$, the revival time generally increases with the mode $n_0$ and hence its energy. The notable exception is for the resonant case when $2n_0/q=1$, where the initial mode couples directly with its degenerate mode, in which case, the revival time remains surprisingly constant regardless of the specific initial mode, $n_0$ in consideration.

\begin{figure}[t]
\centering
%\vspace{-0.2\linewidth}
\includegraphics[width=\columnwidth]{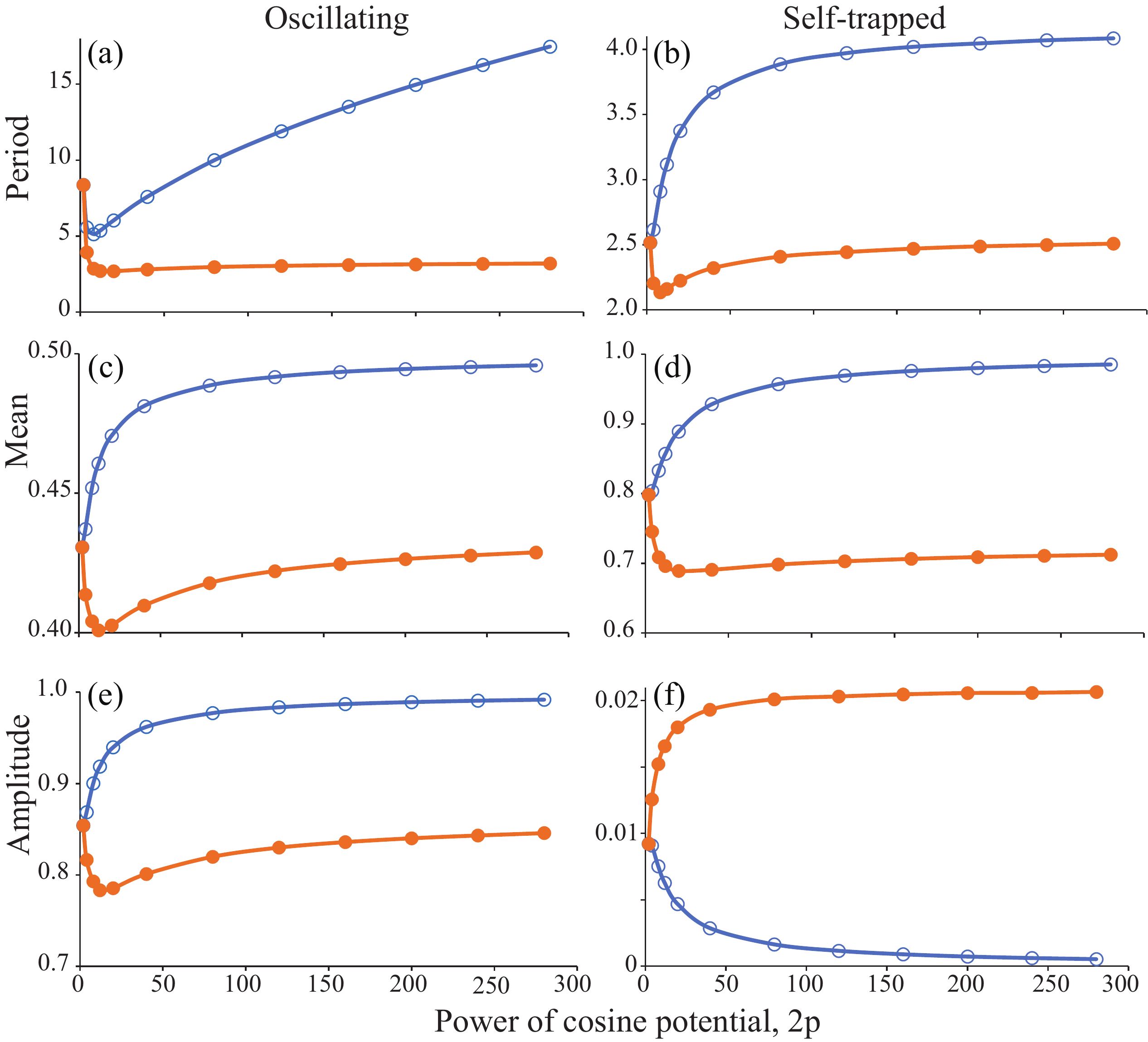}
\caption{(Color online) The influence of the power $2p$ of the cosine potential on the dynamics is represented by tracking the period, mean-value and amplitude of the oscillations/wiggles in the time-evolution of the initial mode.  Left panels are for an oscillating case with $n_0=2,q=2$ and the right panels for self-trapped case with $n_0=2,q=3$. The full circles interpolated by orange lines keep the area under the potential fixed by adjusting the lattice strength $U_0$ while the empty circles interpolated by blue lines keep $U_0$ constant.}
\label{Fig8-Power}
\end{figure}

\subsection{Exponent of the Potential}

We have presented all our results so far for the power of cosine in the lattice potential in Eq.~(\ref{potential}) at $2p=2$. We now show that the conclusions are qualitatively unchanged by that constraint.  Increasing the exponent $2p$ has the effect of narrowing the width of the peaks of the potential with no effect on the lattice periodicity. We consider two options, (1) keeping the strength $U_0$ fixed and (2) adjusting it so that the integrated area of each peak remains constant. The latter option simulates the approach to a sequence of delta functions in the limit $p\rightarrow \infty$, matching a similar limit of a Kronig-Penney model. Figure~\ref{Fig8-Power} illustrates the effects. Fixed $U_0$ makes the potential weaker with increasing power, so the oscillation period increases, consistent with the proportionality of energy to frequency. However, when $U_0$ is adjusted to maintain the strength of the lattice, the period settles into a stable value.  A sharp initial drop is due to the opening up of new ladder paths with larger step size, implicit in the trigonometric expansion in the second form of the potential in Eq.~(\ref{potential}).

The mean value of the oscillation captures the loss of population from the initial state, averaged over a period. The patterns are similar for both oscillating and self-trapped cases, but with different physical interpretations. For the oscillating case, the mean value is proximate to a value of $1/2$ which corresponds to oscillations with maximum amplitude. At fixed $U_0$ as the potential weakens, it is harder to excite more distant modes so that the system tends towards a full amplitude oscillation between the dominant degenerate pair. This is also reflected in the behavior of the amplitude. The initial sharp boost levels off as the maximum amplitude is approached asymptotically. When the strength of the potential is kept constant by adjusting $U_0$, both the amplitude and the mean remain at values that reflect diminished oscillations of the dominant pair, with population acquired by other modes. A sharp drop at low power is once again due to the opening up of new ladder paths.

In the self-trapped cases, the mean values approach unity and the amplitude of the wiggles approaches zero when the potential weakens with increasing power, both indicative of vanishing effect of the lattice. But, when the potential strength is maintained, the narrowing of the potential causes the mean to stabilize at some value representing some loss of population in the initial state, which is accompanied by the characteristic wiggles that settle into some constant amplitude.

There are two main messages here: (1) If the strength of the potential is maintained, the dynamics becomes generally insensitive to the exponent, consistent with settling into universal behavior as the lattice approaches the delta comb limit. (2) The greatest impact of the exponent is at low values when the emergent new ladder pathways have the highest relative weights compared to the fundamental pathway.

\begin{figure}[t]
\centering
%\vspace{-0.2\linewidth}
\includegraphics[width=\columnwidth]{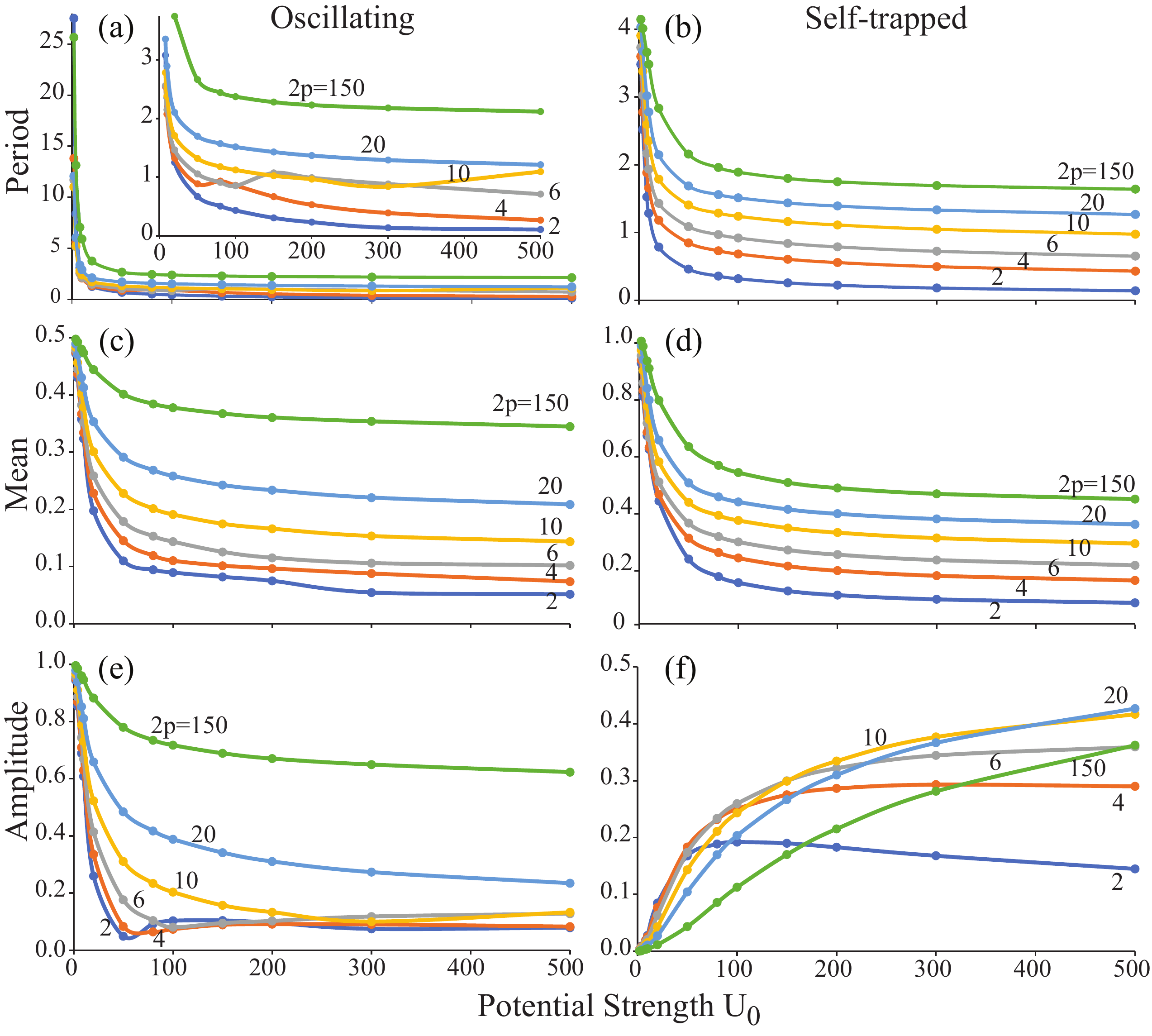}
\caption{(Color online) The influence of the lattice strength $U_0$ of the cosine potential on the dynamics is represented by tracking the period, mean-value and amplitude of the oscillations/wiggles in the time-evolution of the initial mode.  Left panels are for an oscillating case with $n_0=2,q=2$ and the right panels for self-trapped case with $n_0=2,q=3$. The interpolating lines are for guidance only, the multiple lines are for different values of the power of the cosine potential, $2p$.}
\label{Fig9-U0}
\end{figure}

\subsection{Strength of the Potential}

In Fig.~\ref{Fig9-U0}, we present the effect of increasing the strength of potential, fixing  the exponent  $2p$ of the cosine function in the potential Eq.~(\ref{potential}) at different values, to create a picture of the composite effect of the potential strength and width. For both oscillating and self-trapped cases, the increase in the strength of the lattice potential causes a decline in the period of the oscillations consistent with higher energy input as seen in Fig.~\ref{Fig9-U0}. A sharp decline is seen at lower strengths, particularly for the oscillating case, but at higher strengths, the period levels off at values that trend higher with the weakening due to the narrowing of the potential at higher exponents.

The same basic pattern appears for the mean value as well. The decline reflects increased loss of population to modes farther apart on the ladder made energetically accessible by the stronger lattice coupling, but softened by narrowing of the potential at higher exponent.

In the case of strong oscillations, the amplitude acts in tandem with the mean value for the same reasons. However, the amplitude of the wiggles reveal an opposite trend, an increase with $U_0$ at the beginning before leveling off and possibly turning around. This occurs because initially, a stronger potential boosts the coupling between the initial mode $n_0$ and its neighbor $n_0-q$, but as the strength increases further the latter mode can itself couple further along the ladder reducing the amplitude of the initial mode. Notably, the turnaround effect is less pronounced at higher exponents, since the existence of multiple ladder pathways allows the initial mode to directly couple to multiple modes on the ladder, leading to more pronounced oscillations.

The influence of the lattice strength $U_0$ is particularly relevant in the two limits of weak lattice and strong lattice. In the weak lattice limit, clearly there is strong sensitivity to the lattice depth, with even local reversal of trends occurring for small exponents $2p$ that can be attributed to the heightened impact of new coupling pathways opening up.  The strong lattice marks a tight-binding limit, where localized states would be a more appropriate basis and the behavior of individual circulating modes tends to become uniform.

\section{Outlook and Conclusions}
The considerations in this paper indicate some of the most basic experiments that could be done on a ring shaped lattice. The primary criterion for testing the results here is to trap ultracold atoms that maintain coherence in a toroidal trap where an azimuthal lattice structure can be introduced. Confinement of atoms in ring traps has been demonstrated and utilized in experiments for some time now, with LG beams \cite{ramanathan, KENNEDY2014110, Phillips_Campbell_superfluid_2013} as well as via other methods such as with an intensity mask \cite{Campbell_resistive-flow, Phillips_Campbell_hysteresis}. In some of these experiments, sharply focussed blue-detuned lasers have been used to create potential structures along the azimuth, indicating feasibility of a lattice potential with a uniformly spaced multitude. Ring shaped lattices have already been demonstrated by interfering two LG beams with opposite orbital angular momentum (OAM) and offers a particularly convenient method \cite{Padgett, Zambrini:07}. What remains to be done is the clearly feasible step of bringing together these two features, already separately realized, azimuthal lattice potential, and confinement of ultracold atoms in ring traps.

The other necessary feature in our study is the inclusion of rotation. Angular motion of the general frame is clearly an independent consideration that does not impact trapping of atoms in ring shaped lattices. The lattice itself can be easily rotated by introducing a frequency shift between the two constituent LG beams, limited only by the speed of the phase modulation, but rotation periods at the picosecond levels are possible with an electro-optic modulator \cite{Yamane:16}.  Measurement of the quantum states in the ring can be done through various imaging techniques, the phase via mapping to density modulation such as by interference of a circulating condensate with a non-circulating one \cite{Phillips_Campbell_hysteresis, Campbell-current-phase}, and measurements of flow even \emph{in situ} have been demonstratd \cite{Phillips_Stringari-DMD}.

Adiabatic introduction of the lattice can be done in a variety of ways, one possibility in the case of LG beams, would be to use one beam of zero OAM  for confinement, and a pair with opposite  OAM, $\pm\hbar q$ to create the lattice that can be independently controlled, so raising its intensity would adiabatically introduce the lattice. Different values of the OAM would lead to different lattice periodicities in multiples of $q$. Finally, although not absolutely essential, the nonlinearity can be reduced by use of Feshbach resonance to tune atom-atom interactions to approximate the linear behavior \cite{Tiesinga-RMP} assumed here.

The results reported here offer multiple routes for further research, particularly with the introduction of nonlinearity, a subject of our ongoing interest. Specifically, a prior study, co-authored by one of us, showed that rotation sensitivity can be enhanced by use of spin squeezing induced by nonlinearity \cite{Opatrny-Kolar-Das-rotation}. The sensitivity, as observed here, of the amplitude of mode oscillation to rotation could possibly be similarly enhanced, to make it a competitive mechanism for matter wave rotation sensing.

\begin{acknowledgments} We gratefully acknowledge valuable discussions with M. Kol\'{a}\v{r} and D. Schneble, and the support of the NSF under Grant No. PHY-1707878. \end{acknowledgments}

%\bibliography{ring_quantum_optics}

\vfill
\end{document}